\documentclass[api,prl,showpacs,letterpaper,twocolumn,groupedaddress,footinbib,amssymb]{revtex4}
\usepackage[utf8x]{inputenc}
\usepackage[xdvi]{graphicx}
\usepackage{amsmath}
\usepackage{amssymb}
\usepackage{amsthm}
\usepackage{xcolor}

\begin{document}
\begin{abstract}
Spin-polarized scanning tunneling microscopy is identified as a suitable experimental technique to investigate the quantitative quality of Lieb-Robinson bounds on the 
signal velocity. The latest, most general bound is simplified and it is shown that there is a discrepancy by a factor of approximately 4 between the corresponding limit 
speed and some estimated exact velocities in atomic spin chains. The observed discrepancy facilitates conclusions for a further mathematical improvement of 
Lieb-Robinson bounds. The real signal propagation can be modified with several experimental parameters from which the bounds are independent.
This enables the application of Lieb-Robinson bounds as upper limits on the enhancement of the real signal speed for information transport in spintronic devices.
\end{abstract}
  
\pacs{03.65.Fd,75.10.Pq,75.10.Jm,03.67.Hk}

\title{Towards experimental tests and applications of Lieb-Robinson bounds}

\author{K.~Them}
\email[\textit{Electronic mail: }]
    {kthem@physnet.uni-hamburg.de}
\affiliation{Institute of Applied Physics and Microstructure
    Research Center, University of Hamburg,
    Jungiusstr.~11, 20355 Hamburg, Germany}

\date{\today}
\maketitle
\section{Introduction}
In spin-based nanotechnology, a fast information transfer via spin chains is crucial for the efficiency of future
spintronic devices \cite{Khajetoorians27052011,PhysRevLett.108.197204}, which are investigated by means of spin-polarized scanning tunneling microscopy (SP-STM) 
\cite{RevModPhys.81.1495}. There are different experimental parameters which can be used to modify the information transport in magnetic quantum systems, e.g., 
external magnetic fields, differently prepared states, the temperature or the bias voltage for the atom's
magnetization switching. 
The enhancement of the signal velocity is of primary interest.
This naturally raises the question of a possible maximum velocity in quantum spin systems, 
in analogy to the speed of light as maximum velocity in relativity theory. 

In 1972, Lieb and Robinson provided a general limit on signal velocities in
the theory of quantum spin systems, which is expressed in a mathematical bound \cite{LiebRobinsonBound}. 
While the speed of light as maximum velocity is physically realized in nature by photons, it is an interesting question how close
real velocities in magnetic quantum systems can come to the mathematical Lieb-Robinson bound.
There were several investigations of this bound in the 
algebraic framework of mathematical physics \cite{BrunoI,BrunoII,BrunoIII}.
A central aim of these investigations is the improvement of the bound, which means a reduction of the corresponding upper mathematical limit down to 
real existing velocities. However, Lieb-Robinson bounds were never applied to real physical systems and the discrepancy between the bound and exact velocities 
remained unclear. Therefore, the quantitative quality and the corresponding potential for a further improvement of the bounds remained unknown. 

In this paper we suggest that SP-STM is a suitable experimental setup for a quantitative investigation of Lieb-Robinson bounds. 
The latest, most general bound \cite{BrunoII} is presented in a simplified formula \cite{FredenTalk} and 
applied to realistic interactions of magnetic quantum systems as used in 
spin-based nanotechnology \cite{ScienceCover}. It is shown, that this bound, $\mathfrak{B}$, is better by a factor of 100 
than the old bound \cite{BratteliRobinson}  
and approximately 4 times faster than some estimates of exact signal velocities
in atomic spin chains.    
  
The observed discrepancy between the bounds and the exact signal velocities
is investigated. This allows a quantitative analysis for further improvements of Lieb-Robinson bounds and provides an upper limit on 
the enhancement of exact velocities, if experimental parameters are changed which leave the bound invariant.
It is shown that the latest, most general bound derived in \cite{BrunoII} is already in the correct order of magnitude. 
A specific bound which is only valid for the XY-model
\cite{PhysRevLett.99.167201} is incorporated into the discussion. 
A bound $\tilde{\mathfrak{B}}$ is derived which provides a mathematical relation between the bound $\mathfrak{B}$ and the exact velocities.
The interaction between the magnetic STM tip and the sample is identified as an interesting 
experimental parameter which influences the exact signal propagation and $\tilde{\mathfrak{B}}$, but leave $\mathfrak{B}$ invariant.
A corresponding SP-STM experiment is simulated with explicit calculations. 
Algebraic and the numerical method of exact diagonalization are connected with interesting applications to SP-STM.

\section{Magnetic interactions, Lieb-Robinson bounds and exact velocities}
\begin{figure}[h]
\includegraphics*[scale=0.8,viewport=147 503 388 669]{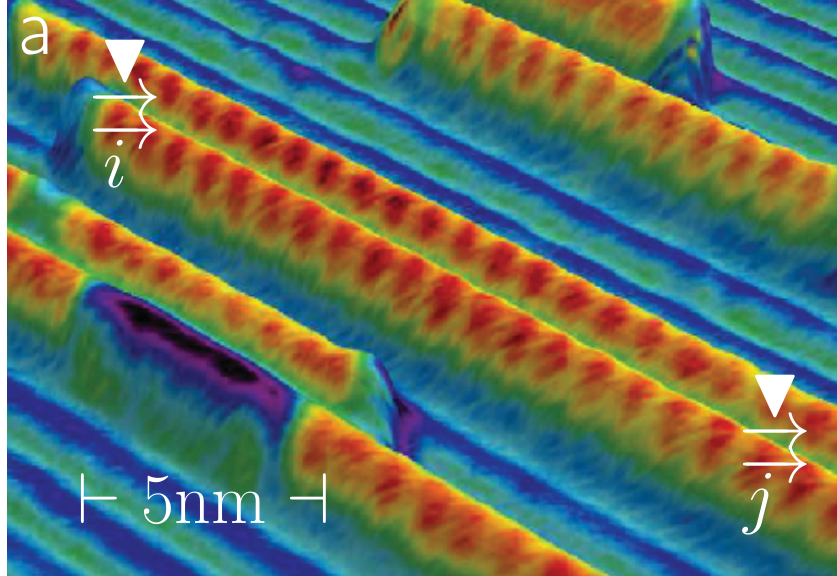}
\includegraphics*[scale=0.92,viewport=146 540 354 675]{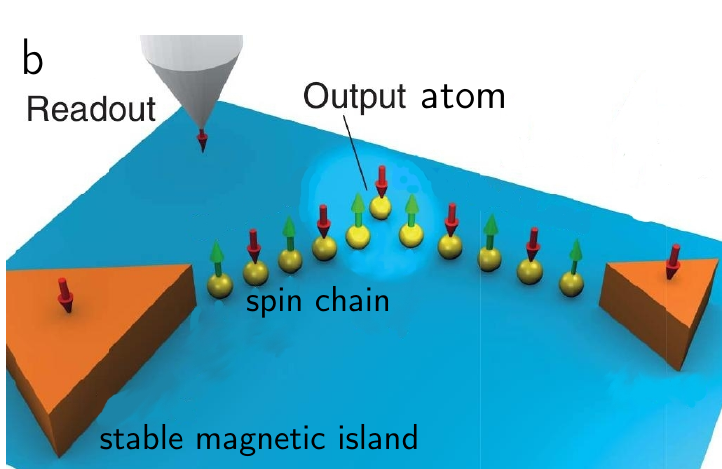}
\caption{Examples of SP-STM experiments, where information is transported via spin chains. a) Iron chains on the surface of an iridium 
crystal \cite{PhysRevLett.108.197204}. A magnetic tip (upper white triangle) starts to act on lattice
site $i$ at time $t=0$. After some time $t>0$, the signal can be detected at lattice site $j$. b) 
An all-spin-based atomic-scale logic device consisting of iron and cobalt atoms
placed on copper(111) \cite{Khajetoorians27052011}. If the magnetization of a triangular island is switched, a signal starts to propagate through the chain 
towards the output atom.}
\end{figure}
Fig. 1 shows two visualized SP-STM experiments, where information is transferred via spin chains consisting of iron atoms on the surface of a) an 
iridium crystal \cite{PhysRevLett.108.197204} and b) copper(111) \cite{Khajetoorians27052011}. Our model calculations will be performed for systems where the 
interaction between the atoms can be 
described by the Heisenberg interaction, which is the case for the atomic-scale spin-based logic device in fig. 1 b).

Magnetic atoms placed on a substrate form a spin system on a lattice $L$. The positions of the magnetic atoms are described by points $i\in L$.
The interaction between the magnetic atoms is often described by the Heisenberg interaction
\begin{equation}\label{eq:Heisenberg}
\Phi_{\mathrm{H}}(\{i,j\})=\sum_{\alpha=\mathrm{x},\mathrm{y},\mathrm{z}}J^\alpha_{ij}S^\alpha_iS^\alpha_j,
\end{equation} where $(S^\alpha_i)$ denotes the $\alpha=\mathrm{x},\mathrm{y},\mathrm{z}$-component of the spin operator at lattice site $i\in L$.
A metallic substrate induces the anisotropy energy
\begin{equation}\label{eq:ani}
\Phi_{\mathrm{ani}}(\{i\})= K(S^\mathrm{z}_i)^2,
\end{equation} where $K$ is a constant.
The action of an external magnetic field is described by
\begin{equation}\label{eq:ExternalField}
\Phi_{\mathrm{B}}(\{i\})= g\mu_\mathrm{B}\vec{B}\vec{S}_i,
\end{equation} where $g$ is a gyromagnetic constant, $\mu_\mathrm{B}$ the Bohr magneton and $\vec{B}$ the external magnetic field.
$\Phi_{\mathrm{ani}}$ and $\Phi_{\mathrm{B}}$ are typical one-body interactions, i.e.,
they act on each single spin.

Magnetic STM tips 
are experimentally used to investigate the spin system and if a measurement is started, a tunneling current starts to flow between the tip and the magnetic atoms.
The tunneling current perturbs the spin system locally and the corresponding interaction is often described by the modified Tersoff-Hamann 
model \cite{PhysRevB.31.805,PhysRevLett.107.027203},
\begin{equation}\label{eq:Tersoff}
H_{\mathrm{tip}}(\{i\})=gI_0\mathcal{P}e^{-2\kappa\sqrt{(i-i_0)^2+h^2}}\vec{m}_{\mathrm{tip}}\cdot\vec{S}_{i},
\end{equation} where $g$ is a coupling constant, $I_0$ is the spin-polarized current averaged over the surface and $\mathcal{P}$ denotes the polarization of the
tunneling current. $\vec{m}_{\mathrm{tip}}$ is a unit vector in the direction of the tip-magnetization. The position of the tip
is given by the height $h$ above the sample and the lattice point $i_0$. $H_{\mathrm{tip}}(\{i_0\})$ is the theoretical idealization, that the tip acts only on the 
atom which is directly under the tip, which is sufficient for our interest. 

In accordance to \cite{Them}, we will distinguish between a \textit{free} spin system (the tip is moved away from the surface such that there is no interaction 
between the tip and the sample) and a \textit{perturbed} spin system, 
which means that there is an interaction between the magnetic tip and the sample.
The Hamiltonian of a subsystem $\Lambda\subset L$ of the free spin system is given by the sum of all 
interactions, described in eq. (\ref{eq:Heisenberg}) - (\ref{eq:ExternalField}), which are contained in $\Lambda$:
\begin{equation}
H_\Phi(\Lambda)=\sum_{X\subset\Lambda}\Phi(X).
\end{equation} Thus, the Hamiltonian of the spin system switches from 
$H(\Lambda)$ to $H(\Lambda)+H_{\mathrm{tip}}(\{i_0\})$ in the moment ($t=0$), when the tunneling current starts to interact with the magnetic atoms.  
The action of the magnetic tip influences the dynamics of the spin system and disturbances start to propagate through the chain.
In the following text we will use the notation $P\equiv H_{\mathrm{tip}}(\{i_0\})$ to emphasize the mathematical connection to \cite{BratteliRobinson}.
The dynamics of the free spin system is given by the Heisenberg relations
\begin{equation}\label{eq:Dynamic}
 S^\mathrm{z}_i(t)=e^{i\frac{t}{\hbar}H_\Phi(\Lambda)}S^\mathrm{z}_ie^{-i\frac{t}{\hbar}H_\Phi(\Lambda)}=\tau^\Lambda_t(S^\mathrm{z}_i).
\end{equation} 
The corresponding dynamics of the perturbed system is given by
\begin{equation}\label{eq:DynamicP}
\tau^{\Lambda P}_t(S_i^\mathrm{z})=e^{i\frac{t}{\hbar}(H_\Phi(\Lambda)+P)}S_i^\mathrm{z}e^{-i\frac{t}{\hbar}(H_\Phi(\Lambda)+P)}.
\end{equation}  
In thermal equilibrium, the state of the free (finite dimensional) system is given by the Gibbs state $\omega^\beta_\Lambda$ whose expectation value is given by
\begin{equation}
 \omega^\beta_\Lambda(S^z_i)=\frac{Tr(e^{-\beta H_\Phi(\Lambda)}S^z_i)}{Tr(e^{-\beta H_\Phi(\Lambda)})},
\end{equation} where $\beta$ is the inverse temperature.
The dynamics for the exact velocities is now given by
\begin{equation}\label{eq:ExactVelocity}
 \omega^\beta_\Lambda(\tau^{\Lambda P}_t(S^z_j))\equiv\langle S^\mathrm{z}_j\rangle(t),
\end{equation} where the place $i_0$ of the perturbation $P$ is different to the place $j$ of the observed spin operator $S^\mathrm{z}_j$, e.g., the output atom in 
fig. 1 b).

If the observables $S^\mathrm{z}_j$ and $S^\mathrm{z}_i(t)$ commute, i.e.,
\begin{equation}
\parallel[S^\mathrm{z}_i(t),S^\mathrm{z}_j]\parallel=0,
\end{equation} it is implied that no signal can propagate from the lattice site $i$ to the site $j$ within the time $t$.
A Lieb-Robinson bound provides an upper bound on this commutator in the slightly more general form
\begin{equation}\label{eq:StartingPoint}
C_A(t,X)=\sup_{B\in\mathfrak{A}_X;\parallel B\parallel=1}\parallel [B,\tau_t(A)]\parallel,
\end{equation} where $A\in\mathfrak{A}_Y$ and $\mathfrak{A}_Y$ is the $C^*$-algebra associated with $Y\subset L$. 
$C_A(t,X)$ is a quantitative measure for the amount of information, which is propagated from $Y$ to $X$ at time $t$. 
If this number is close to zero and much smaller than the norm of $A$ for all times $t$, with $0<t<t'$, then there can be no significant information transport from $Y$ 
to $X$ within the time $t'$. The old Lieb-Robinson bound in \cite{BratteliRobinson}
(Theorem 6.2.11.), is given by:
\begin{equation}\label{eq:Commutator}
 \parallel[S^\mathrm{z}_x(t),S^\mathrm{z}_y]\parallel\leq 2\parallel S^\mathrm{z}_x\parallel \parallel S^\mathrm{z}_y\parallel
 e^{-\mid t\mid(\xi \mid x-y\mid/\mid t\mid-2\parallel \Phi \parallel_\xi)}\dot{=}\mathfrak{L}(t),
\end{equation} where the interaction norm is given by
\begin{equation}\label{eq:InteractionNorn}
 \parallel\Phi\parallel_\xi=\sup_{x\in L}\sum_{X\ni x}|X|(2s+1)^{2|X|}e^{\xi D(X)}\parallel\Phi(X)\parallel<+\infty
\end{equation} for some $\xi>0$, where $|X|$ is the number of points in $X$, $D(X)$ is the diameter of $X$ and $s$ is the spin quantum number.

Now we will present a simplified formula \cite{FredenTalk} for the latest, most bound \cite{BrunoII}. 
The mathematical proof is presented in the appendix.
Eq. (\ref{eq:StartingPoint}) and (\ref{eq:StartingPoint2}), 
the next steps and the intermediate result eq. (\ref{eq:EQCopy}) are taken from \cite{BrunoII}, but our iteration \cite{FredenTalk} provides:
\begin{equation}\label{eq:NewBound}
C_A(t,X)\leq 2\parallel A \parallel \sum_{\gamma; X\rightarrow Y}\frac{|2t|^{L(\gamma)}}{L(\gamma)!}w(\gamma)\dot{=}\mathfrak{B}(t),
\end{equation} where $L(\gamma)$ is the length of the path $\gamma$ from $X$ to $Y$ and $w(\gamma)$ is the weight of the path, defined 
in eq. (\ref{eq:wight}). The proof states, that only interactions on the sets $Z\in\partial X$ of the boundary of $X$, eq. (\ref{eq:boundary}), 
contribute and the bound is clearly independent of one-body interactions, e.g., external magnetic fields, anisotropy energies and the modified Tersoff-Hamann model.
In a lattice of dimension $d$ the number of paths with length $L$ is bounded by $(2(2d-1))^L$.
For nearest neighbor Heisenberg interaction the bound $\mathfrak{B}$ can be simplified to \cite{FredenTalk}
\begin{equation}\label{eq:NewBound1D}
\parallel [S^{\mathrm{z}}_i(t),S^{\mathrm{z}}_j(t')] \parallel\leq s^2 \bigg( \frac{v|t-t'|}{|i-j|} \bigg)^{|i-j|},
\end{equation} with $v=4e(2d-1) Js^2$, where $d$ is the dimension of the lattice and $s$ is the spin quantum number.
In contrast to the old bound $\mathfrak{L}$ \cite{BratteliRobinson}, the new bound $\mathfrak{B}$ is independent of the arbitrary choice of a number 
$\xi>0$ and there is no 
factor $|X|(2s+1)^{2|X|}e^{\xi D(X)}$ which provides an unnecessary increase of the corresponding limit speed. 
The specific bound in \cite{PhysRevLett.99.167201} for the XY-model is 
multiplied with the square $n^2$ of the chain length $n$, which is certainly a disadvantage for a large chain length and from which the general bounds are independent. 

However, there is no mathematical relation between the exact velocities, eq. (\ref{eq:ExactVelocity}), and the bound $\mathfrak{B}$. 
To prevent possible errors which might arise from a physical interpretation without a
mathematical relation, we will also introduce a strictly mathematically 
related bound $\tilde{\mathfrak{B}}$ on the exact velocities. This is done with the help
of Proposition 5.4.1. in \cite{BratteliRobinson} and one obtains
\begin{align}\label{eq:RelatedBound}
 \mid\omega(\tau^P_t(S^\mathrm{z}_i))-\omega(S^\mathrm{z}_i)\mid \leq & \sum_{n=1}^\infty\int_0^tdt_1\cdots\int^{n-1}_0dt_n \times \notag\\
& \times(2\parallel P\parallel)^{n-1}\mathfrak{B}(t_n)
\parallel S^\mathrm{z}_i\parallel \notag\\ 
=& \parallel S^\mathrm{z}_i\parallel\int^t_0dt'\mathfrak{B}(t')\times \notag\\
&\times\sum^\infty_{n=1}\frac{(t-t')^{n-1}}{(n-1)!}(2\parallel P\parallel)^{n-1} \notag\\
=& \parallel S^\mathrm{z}_i\parallel \frac{d}{d\lambda}\frac{e^{\lambda t}-1}{\lambda}\dot{=}\tilde{\mathfrak{B}}(\parallel P\parallel,t),
\end{align} for $\lambda=2\parallel P\parallel$ after the differentiation in the last line. 
The bound $\tilde{\mathfrak{B}}$ depends on the norm of the perturbation, 
$\parallel P\parallel$, which contains the parameters of the modified Tersoff-Hamann model. 
Thus, we have prevented potential errors which might arise from a physical interpretation by the 
cost of an additional dependence of the parameters from the modified Tersoff-Hamann model.
But still, the bound is independent of the other one-body interactions and valid for all temperatures.

\section{Numerical calculations and information transfer in spin chains}

\begin{figure}[h]
\includegraphics*[scale=1.05,viewport=150 540 354 678]{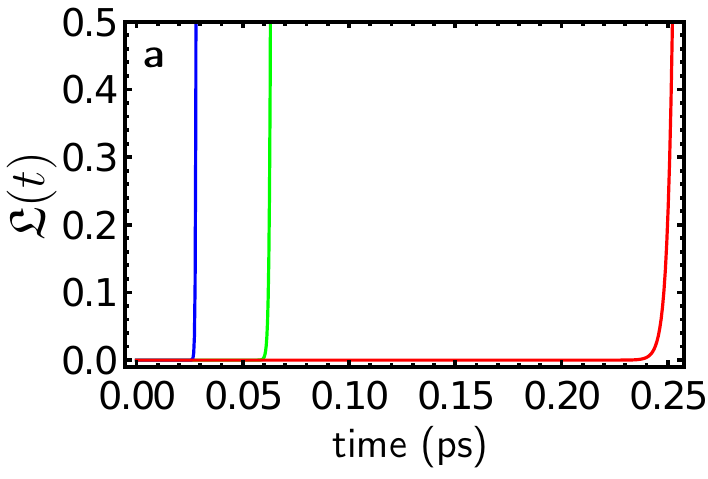}
\includegraphics*[scale=1.03,viewport=150 535 354 675]{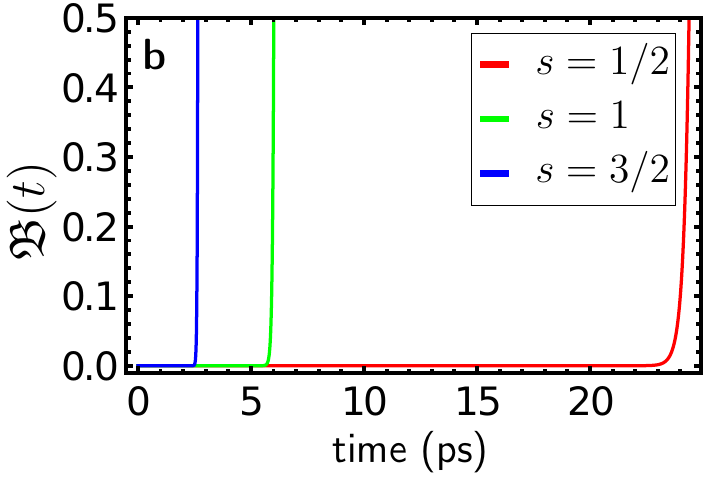}
\caption{The old a) and the new bound b) are compared for the $s=1/2,1$ and $3/2$ nearest neighbor $J=1$ meV Heisenberg chain consisting of 100 quantum spins. 
The velocity of the new bound is approximately 100 times slower and therefore 100 times closer to real existing velocities.
}
\end{figure}
First we will solve the old bound $\mathfrak{L}$ and the new bound $\mathfrak{B}$ for the interactions eq. (\ref{eq:Heisenberg})-(\ref{eq:ExternalField}), compare 
them as a function of time and discuss the observed discrepancy. We will then solve the bound $\tilde{\mathfrak{B}}$ to compare the latest, most general bound 
$\mathfrak{B}$ and the exact signal velocities, eq. (\ref{eq:ExactVelocity}), which are generated in the simulated SP-STM experiment.

Fig. 1 shows a) the original bound $\mathfrak{L}$ and b) our simplified bound $\mathfrak{B}$ for the $s=1/2,1$ and $3/2$ 
nearest neighbor Heisenberg quantum spin chain with interaction strength $J=1$ meV.
The chain length is chosen to be $l=100$ lattice sites. 
While the old bound $\mathfrak{L}(t)$ states for $s=1/2$ that it is impossible to transfer 
information through the chain faster than $0.24$ ps, the new and better bound $\mathfrak{B}(t)$ states that it is impossible to do this faster than $23.5$ ps. 
The new bound is therefore approximately 100 times slower, i.e., stays longer close to zero with increasing time, and is 100 times closer to real existing 
velocities. In the original bound $\mathfrak{L}$ the time is multiplied with the factor $|X|(2s+1)^{2|X|}e^{\xi D(X)}$, from which the new bound $\mathfrak{B}$ is 
independent. This factor provides an unnecessary increase of the corresponding limit speed. In our calculations we have $|X|(2s+1)^{2|X|}e^{\xi D(X)}\approx86,99$ and 
$\xi$ was chosen nearly to $1$, such that $\mathfrak{L}(t)$ has a minimum value. 
Thus, the disappearance of this factor in the new bound $\mathfrak{B}$ provides the main contribution of the 100 fold improvement. 

Next we estimate some exact velocities in spin chains and include the interaction of the magnetic tip. 
A $s=1$ nearest neighbor interaction Heisenberg quantum spin chain consisting of 8 atoms with 
anisotropy energy eq. (\ref{eq:ani}) will be used for our next example:
\begin{equation}\label{eq:Hamiltonian}
H(\Lambda)=\sum_{i,i+1}^8J\vec{S}_i\cdot\vec{S}_{i+1}+\sum_{i=1}^8K(S^{\mathrm{z}}_i)^2,
\end{equation} where we choose $J=1$ meV and $K=2$ meV, corresponding to typical values of magnetic atoms on the surface of a metallic substrate.
While the particle number is mostly the limiting factor for numerical methods, Lieb-Robinson bounds can be solved for an arbitrary
number of quantum spin particles.

Fig. 3 a) shows a comparison between three different exact velocities (colored lines) and the upper bound $\mathfrak{B}(t)$ (black line).
The parameter of the modified Tersoff-Hamann model are chosen to be
\begin{equation}
 \parallel P\parallel= gI_0\mathcal{P}e^{-2\kappa\sqrt{h^2}}=1,2\,\mathrm{and}\,4\,\mathrm{meV}.
\end{equation}
The limit speed, obtained from $\mathfrak{B}(t)$, is approximately four times faster than the exact velocities and independent of $\parallel P\parallel$. 
This means that there is no possibility to enhance the exact velocity above the black line, by changing the experimental parameters  
temperature, external magnetic fields, anisotropy energies, differently prepared states or the variables $gI_0\mathcal{P}e^{-2\kappa\sqrt{h^2}}$ in 
$\parallel P\parallel$. However, this comparison between the bound $\mathfrak{B}$ and the exact velocities is based on a physical interpretation instead of a 
mathematical relation. Therefore, we will solve the bound $\tilde{\mathfrak{B}}$, eq. (\ref{eq:RelatedBound}),which relates $\mathfrak{B}$ and the exact signal velocities, eq. 
\ref{eq:ExactVelocity}, mathematically. $\tilde{\mathfrak{B}}$ contains a dependence of the perturbation $\parallel P\parallel=gI_0\mathcal{P}e^{-2\kappa\sqrt{h^2}}$,
but is still independent of the other experimental parameters mentioned above.

\begin{figure}[h]
\includegraphics*[scale=1.05,viewport=148 532 354 677]{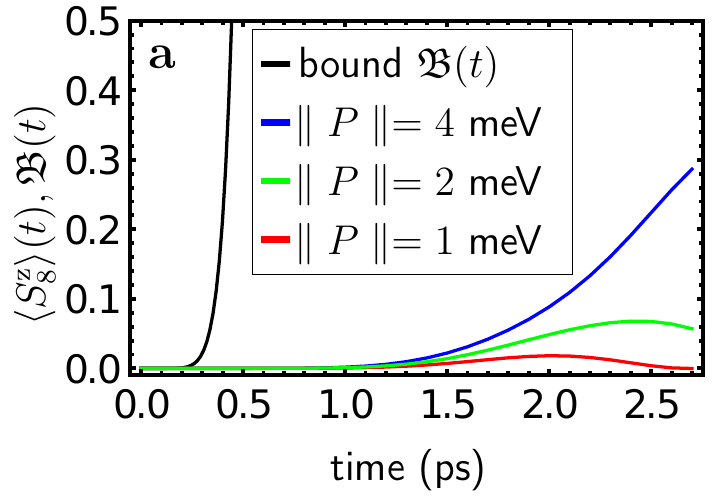}
\includegraphics*[scale=1.07,viewport=151 532 358 675]{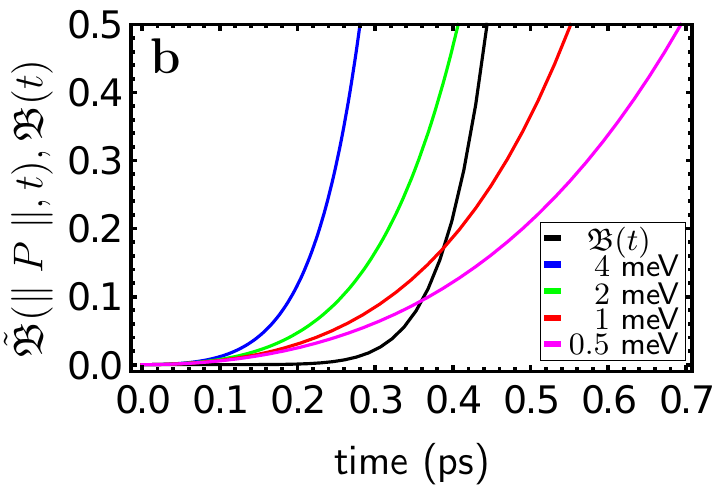}
\caption{a) The bound $\mathfrak{B}(t)$ (black line) is approximately 4 times faster than the exact velocities (colored lines), for the case of
a Heisenberg quantum spin chain with $J=1$ meV and a length of $8$ lattice sites. b)
The bound $\mathfrak{B}(t)$ (black line) for the physical interpretation is compared with the bounds $\tilde{\mathfrak{B}}(\parallel P\parallel,t)$ 
(colored lines) which give a strictly mathematically related bound on the exact velocities. 
The interaction strengths of the modified Tersoff-Hamann model are chosen to be $\parallel P\parallel=4,2,1$ and $0.5$ meV.
}
\end{figure}
Fig. 3 b) shows a comparison of the improved bound $\mathfrak{B}(t)$ and the bounds 
$\tilde{\mathfrak{B}}(\parallel P\parallel,t)$, which give a strictly mathematically related bound on the exact velocities, but depend on the Tersoff-Hamann model.
We have used the parameters $\parallel P\parallel=0.5,1,2\,\mathrm{and}\,4\,\mathrm{meV}$
for the bound $\tilde{\mathfrak{B}}$, as in the case of the exact velocities.
It can be seen that the bounds $\tilde{\mathfrak{B}}(\parallel P\parallel,t)$ rise a little bit earlier than $\mathfrak{B}(t)$, but for low energies the increase
of $\tilde{\mathfrak{B}}(\parallel P\parallel,t)$ comes more and more slower. However, $\tilde{\mathfrak{B}}$ and $\mathfrak{B}$ are basically of the same order.

There are different reasons which are responsible for the discrepancy by a factor of approximately 4.
The generality and the way of the mathematical construction of the bounds are two of these reasons. 
There are two kinds of generalities. The first generality, mentioned by the authors in \cite{PhysRevLett.99.167201}, 
is the applicability to a large class of model systems. They used this reason as motivation for the derivation of the specific bound \cite{PhysRevLett.99.167201}, 
which is only valid for the XY-model. The second kind of generality is the
independence of one-body interactions and the validity for all states.
Since we have chosen the numerical method of exact diagonalization, we are able to treat the
problem quite realistically and the values which are obtained from the calculation are exact in the framework of the chosen Hamiltonian. 
However, we have to choose certain parameters of the Hamiltonian which change the values of the 
exact speed but leave the bounds invariant. Furthermore, a specific state has to be chosen for the exact numerical calculation while the bound is valid 
for all states. Therefore, there is some freedom of choice on the side of the numerical calculation, which is not on the side of the bounds.
The reflections while the signal propagates through the chain might be described by some kind of a random walk. 
These reflections depend on the choice of the state. Therefore, one might reduce the reflections in the signal propagation by choosing a different initial state and 
enhance the exact speed.

\section{Conclusion} 
SP-STM was identified as a suitable experimental setup for a quantitative investigation of Lieb-Robinson bounds.
On fig. 2 it was shown that a simplified version of the latest, most general bound, $\mathfrak{B}$, is better by a factor 100 than the old bound $\mathfrak{L}$.
This is mainly caused by the disappearance of the factor $|X|(2s+1)^{2|X|}e^{\xi D(X)}$ in $\mathfrak{B}$.
$\tilde{\mathfrak{B}}$ was derived to provide a strictly mathematical relation between Lieb-Robinson bounds and the exact signal velocities in the simulated experiment. 
On fig. 3 it was shown that exact signal velocities in spin chains are approximately 4 slower than the limit speed provided by $\mathfrak{B}$.
This analysis provided the result that the bound $\mathfrak{B}$ is already in the correct order of magnitude in view 
to exact velocities occurring in realistic magnetic quantum systems. 

The modification and enhancement of the signal velocity by a change of experimental parameters were 
investigated, because of the applications to spintronic devices. 
It was shown that
external magnetic fields, the temperature, anisotropy energies and differently prepared states cannot increase the exact, realistic signal velocity, in fig. 3, 
more than a factor of 4, because the limit speed is independent of these parameters.
The parameters of the modified Tersoff-Hamann model change the output signals of the last chain atom and the bound $\tilde{\mathfrak{B}}$, but 
leave $\mathfrak{B}$ invariant. 
The derivation of better, eventually specific, bounds and the use of several different experimental parameters for the exact velocities would give a further reduction 
of the discrepancy. This future work connects algebraic and numerical methods with interesting applications to experimental realizations.

\acknowledgments
I would like to thank Klaus Fredenhagen for the derivation of eq. (\ref{eq:NewBound}) and (\ref{eq:NewBound1D}) and Thomas Hack, Alex Khajetoorians, Andr\'{e} Kubetzka, 
Roberto Mozara, Elena Y. Vedmedenko, Jens Wiebe and R. Wiesendanger for interesting discussions.
Support by the DFG (SFB 668) and by the ERC Advanced Grant "FURORE" is gratefully acknowledged.

\bibliography{ArticlesImprovedBound}
\section{Appendix. Proof for the simplified formula}
Now we will state the proof for our bound $\mathfrak{B}$, in which the starting points eq. (\ref{eq:StartingPoint}) and (\ref{eq:StartingPoint2}), 
the next steps and the intermediate result eq. (\ref{eq:EQCopy}) are taken from \cite{BrunoII}.
We assume that each algebra $\mathfrak{A}_X$ has a time evolution as a strongly continuous one-parameter group of $^*$-automorphisms,
which is checked by Theorem 6.2.4. in \cite{BratteliRobinson}. 
There is an integral equation for the full time evolution $\tau_t$ of $A\in\mathfrak{A}_\Lambda$:
\begin{equation}
\tau_t(A)=A+i \sum_{X\cap\Lambda\neq\emptyset}\int^t_0 dt'\tau_{t'}([\Phi(X),\tau^{\mathrm{ultra}}_{t-t'}(A)]), 
\end{equation} where $\tau^{\mathrm{ultra}}_t(\mathfrak{A}_\Lambda) \subset\mathfrak{A}_\Lambda$.
An iteration provides a solution under rather general conditions on the interaction.
We have
\begin{align}\label{eq:StartingPoint}
C_A(t,X)=&\sup_{B\in\mathfrak{A}_X;\parallel B\parallel=1}\parallel [B,\tau_t(A)]\parallel\notag \\
=&\sup_{B\in\mathfrak{A}_X;\parallel B\parallel=1}\parallel [\tau_{-t}\tau_t^X(B),A]\parallel.
\end{align} 
There is a differential equation for $f(t)=[\tau_{-t}\tau_t^X(B),A]$:
\begin{equation}\label{eq:StartingPoint2}
\frac{1}{i}\frac{d}{dt}f(t)=\sum_{Z\in\partial X} [[\tau_{-t}(\Phi(Z)),\tau_{-t}\tau_t^X(B)],A],
\end{equation} where $\partial X$ is the boundary of $X$, which is given by
\begin{equation}\label{eq:boundary}
\partial X=\{Z\subset L;Z\cap X\neq \emptyset, Z \nsubseteq X \}.
\end{equation} The Jacobi identity for double commutators states
\begin{align}
[[\tau_{-t}(\Phi(Z)),\tau_{-t}\tau_t^X(B)],A]&=[[\tau_{-t}(\Phi(Z)),A],\tau_{-t}\tau_t^X(B)]\notag \\
&+[\tau_{-t}(\Phi(Z)),\underbrace{[\tau_{-t}\tau_t^X(B),A]}_{f(t)}],
\end{align} and we define
\begin{equation}
 a(t)\dot{=}\sum_{Z\in\partial X}[[\tau_{-t}(\Phi(Z)),A],\tau_{-t}\tau_t^X(B)].
\end{equation} The differential equation for $f$ can now be expressed as
\begin{equation}
 \frac{1}{i}\frac{d}{dt}f(t)=a(t)+[H(t),f(t)].
\end{equation} We have
\begin{equation}
 \frac{1}{i}\frac{d}{dt}U(t)=H(t)U(t), \quad U(0)=1
\end{equation} and define
\begin{equation}
 g(t)\dot{=}U^{-1}(t)f(t)U(t).
\end{equation} A relation between $g$ and $a$ is obtained by
\begin{align}
 \frac{1}{i}\frac{d}{dt}g(t)=&U^{-1}(t)\frac{1}{i}\frac{d}{dt}f(t)U(t)\notag \\
 &-U^{-1}(t)[H(t),f(t)]U(t)\notag \\
 =&U^{-1}(t)a(t)U(t).
\end{align} It follows that
\begin{equation}
 g(t)=g(0)+\int^t_0 dsU^{-1}(s)a(s)U(s)
\end{equation} and
\begin{equation}
 f(t)=f(0)+U(t)\int^t_0 dsU^{-1}(s)a(s)U(s)U^{-1}(t).
\end{equation}
Now we can state an inequality for $C_A(t,X)$, which crucially depend on $a$:
\begin{equation}\label{eq:EQCopy}
C_A(t,X)\le C_A(0,X)+2\sum_{Z\in\partial X}\int^{|t|}_0ds\,C_A(s,Z)\parallel\Phi(Z) \parallel.
\end{equation} The iteration involves finite sequences of sets $Z_1,...,Z_n$ with
\begin{equation}
Z_1\in \partial X, Z_2\in\partial Z_1,...,Z_n\in\partial Z_{n-1}, \quad Z_n\cap Y\neq\emptyset,
\end{equation} for $A\in\mathfrak{A}_Y$. Such a sequence is called a path $\gamma$ of length $L(\gamma)=n$ from $X$ to $Y$.
The weight of a path is defined by
\begin{equation}\label{eq:wight}
w(\gamma)\dot{=}\prod^n_{i=1}\parallel \Phi(Z_i)\parallel,
\end{equation} which enables us to estimate the simplified formula for the bound eq. (\ref{eq:NewBound}) and eq. (\ref{eq:NewBound1D}).

\end{document}